\documentclass[sigplan,10pt]{acmart}
\renewcommand\footnotetextcopyrightpermission[1]{}
\usepackage{fancyhdr}
\usepackage[normalem]{ulem}
\usepackage{microtype}
\usepackage{algorithm}
\usepackage{algpseudocode}
\usepackage[english]{babel}         
\usepackage[utf8]{inputenc}
\usepackage[T1]{fontenc}
\usepackage{colortbl}
\usepackage{multirow}
\usepackage{booktabs}
\usepackage{times}
\usepackage{tikz,pgfplots}
\usetikzlibrary{pgfplots.groupplots}
\usepackage{pgfplotstable}
\usepgfplotslibrary{external} 
\usetikzlibrary{patterns}
\usepgfplotslibrary{fillbetween}
\usetikzlibrary{fadings}
\usepackage{ifthen}
\usepackage{amssymb,amsfonts}
\usepackage{acronym}
\usepackage{pifont}
\usepackage{listings}

\acrodef{HAV}{hardware assisted virtualization}
\acrodef{HPTW}{hardware page table walker}
\acrodef{GPA}{guest physical address}
\acrodefplural{GPA}{guest physical addresses}
\acrodef{HPA}{host physical address}
\acrodef{PT}{page table}
\acrodef{PUD}{page upper directory}
\acrodef{PGD}{page global directory}
\acrodef{VMCS}{virtual machine control structure}

\newcommand\dingnum[1]{\ding{\numexpr181+#1}}

\AtBeginDocument{%
  \providecommand\BibTeX{{%
    \normalfont B\kern-0.5em{\scshape i\kern-0.25em b}\kern-0.8em\TeX}}}

\begin{document}

%
\title{Intel Page Modification Logging, a hardware virtualization feature: study and improvement for virtual machine working set estimation\\
}

%
\author{Stella Bitchebe}
\author{Djob Mvondo}
\author{Alain Tchana}
\affiliation{%
  \institution{Ecole Normale Supérieure de Lyon}
}
\email{prenom.nom@ens-lyon.fr}
\author{Laurent Réveillère}
\affiliation{%
  \institution{Laboratoire Bordelais de Recherche en Informatique}
}

\email{prenom.nom@labri.fr}
\author{Noël de Palma}
\affiliation{%
  \institution{Laboratoire d'Informatique de Grenoble}
}

\email{prenom.nom@lig.fr}
\begin{abstract}
	Intel \textit{Page Modification Logging} (PML) is a novel hardware feature for tracking virtual machine (VM) accessed memory pages.
This  task is essential in today's data centers since it allows, among others, checkpointing, live migration and working set size (WSS) estimation.
Relying on the Xen hypervisor, this paper studies PML from three angles: power consumption, efficiency, and performance impact on user applications.
Our findings are as follows.
First, PML does not incur any power consumption overhead.
Second, PML reduces by up to 10.18\% both VM live migration and checkpointing time.
Third, PML slightly reduces by up to 0.95\% the performance degradation on applications incurred by live migration and checkpointing.
Fourth, PML however does not allow accurate WSS estimation because read accesses are not tracked and hot pages cannot be identified.
A naive extension of PML for addressing these limitations could lead to severe performance degradation (up to 34.8\%) for the VM whose WSS is computed.

This paper presents Page Reference Logging (PRL), a smart extension of PML for allowing both read and write accesses to be tracked.
It does this without impacting user VMs.
The paper also presents a WSS estimation system which leverages PRL and shows how this algorithm can be integrated into a data center which implements memory overcommitment.
We implement PRL and the WSS estimation system under Gem5, a very popular hardware simulator.
The evaluation results validate the accuracy of PRL in the estimation of WSS.
They also show that PRL incurs no performance degradation for user VMs.
\end{abstract}

%
%

%
\settopmatter{printfolios=true}
\maketitle

\section{Introduction}
\label{introduction}

Virtualization has become the foundation of data centers as it allows resource mutualization between multiple clients while ensuring isolation.
Virtualization also provides adequate support for administration including resource management.
In this context, memory page tracking is a key mechanism which allows keeping track of memory accesses by virtual machines, so that the hypervisor can improve memory management for different services it implements. Memory page tracking is at the heart of several essential tasks such as checkpointing~\cite{Zhang:2013:OVC:2535461.2535463} (for recovery after failure), live migration~\cite{Clark:2005:LMV:1251203.1251223} (for maintenance and dynamic packing) and working set size (WSS) estimation~\cite{Denning:1968:WSM:363095.363141} (for memory overcommitment~\cite{Overcommit} and fast restore~\cite{Zhang:2011:FRC:1952682.1952695}).

The widely used approach for implementing memory page tracking relies on present bit invalidation.
Such an approach leads to severe performance degradation (caused by the generated page faults), especially when a significant amount of pages needs to be tracked like in WSS estimation~\cite{Nitu:2018:WSS:3203302.3179422,180153,Waldspurger:2002:MRM:844128.844146,Jones:2006:GMB:1168857.1168861,Lu:2007:VMM:1364385.1364388,Kim:2015:DMP:2818950.2818967}.
We assessed this using a synthetic application which parses an array.
The present bit is invalidated every second for all the VM's memory pages.
We measured up to 96,22\% of performance degradation for a 1GB VM memory size.
To minimize the overhead of this approach, VMware's WSS estimation technique applies invalidation on a reduced (100) random page sample \cite{Waldspurger:2002:MRM:844128.844146}.
But this alternative proved to result in an unprecise WSS estimation~\cite{Nitu:2018:WSS:3203302.3179422}, also assessed in this paper (see Section~\ref{evaluations}).

Intel in collaboration with VMware~\cite{theRegister} started in 2015 to release processors (e.g., Broadwell Xeons) equipped with \textit{Page-Modification Logging} (PML for short)~\cite{pml}, a memory page tracking technology. 
When enabled, the Memory Management Unit (MMU) logs (in a RAM area) any guest physical address (GPA) which leads to the setting of the dirty bit in the extended page table (EPT) during page walks (more details below).

This paper presents two main contributions: (1) a deep analysis and evaluation of the PML technology, and (2) a smart extension of PML which addresses the issues that we identified with PML.

\textit{(First contribution)} Traditionally, new hardware evolutions (especially for Hardware Assisted Virtualization (HAV)) lead to investigations regarding their effectiveness compared with previous solutions.
For instance recently, the introduction of Extended/Nested Page Table~\cite{Uhlig} led to many studies~\cite{Wang2011,Gandhi,Bhargava} comparing it with shadow paging~\cite{Waldspurger:2002:MRM:844128.844146}.
In the same spirit, we propose a deep study of PML from three perspectives: power consumption, efficiency, and performance impact on user applications.
Our findings are summarized as follows.
First, PML incurs no power consumption overhead.
Second, PML reduces both VM live migration and checkpointing time (by up to 10.18\%).
Third, PML slightly reduces (by up to 0.95\%) the performance degradation on applications incurred by live migration and checkpointing, which is not negligible for tail latencies~\cite{tailLatency}.
Fourth, PML however does not allow accurate WSS estimation.

More precisely, the current PML design includes several limitations which prevent effective WSS estimation.
We classify them in two types:
\begin{itemize}
	\item Hard limits, related to PML features: (1) PML does not track every page access.
	It only tracks page modifications, making it inaccurate for read only and mix (read-write) workloads;
	(2) PML does not allow tracking of hot pages.
	It only logs a page once (even if accessed several times), thus cold pages are likely to be counted in the WSS, over-estimating the latter (which leads to memory waste).	
	\item Soft limits, related to PML overhead: PML incurs an unacceptable overhead for the VM whose WSS is estimated.
	This is caused by the fact that PML generates VMExits which are handled by the CPUs of the VM whose WSS is computed.
	The handler of that VMExit can consume a significant CPU time, which is taken from the VM's CPU quota.
	We measured up to 34.8\% of performance degradation when PML is activated and used for WSS estimation.
	This is not acceptable for cloud users because they are not the beneficiaries of WSS estimation.
	WSS estimation is executed on the account of the data center operator.
\end{itemize}
According to these limitations, it appears that Intel developers mainly focused on write workload (hence the name PML), which already represents a significant contribution, but does not allow an effective implementation of WSS estimation.

WSS estimation is an essential task for data center operators because it allows, among others, memory overcommitment~\cite{Nitu:2018:WSS:3203302.3179422,Nitu:2018:WZP:3190508.3190537} (periodically adapting the VM memory size according to its actual needs), fast restore~\cite{Zhang:2011:FRC:1952682.1952695} and efficient processor cache partitioning \cite{intelCatRealTime}.
Regarding memory overcommitment for example, its implementation and adoption are necessary for the following reasons.
First, VM owners use to over-estimate resources~\cite{Delimitrou:2014:QRQ:2541940.2541941,Jyothi:2016:MTA:3026877.3026887} for their tasks.
Jyothi et al.~\cite{Jyothi:2016:MTA:3026877.3026887} analyzed resource reservation for a 50k nodes production data center and found that 75\% of jobs were over-provisioned (even at their peak), with 20\% of them over $\times 10$ over-provisioned.
E. Cortez et al.~\cite{Cortez:2017:RCU:3132747.3132772} made similar observations in recent Microsoft Azure cloud traces.
Second, some types of cloud native workloads fundamentally rely on dynamic management of overcommitted VMs.
In particular, this is the case for serverless/FaaS (Function as a Service) systems, whose design and pricing model are inherently tied to aggressive packing of hundreds or thousands of micro VMs on the same physical machine \cite{brooker18serverless,wang18peeking}. 
Third, memory is a limited resource whose evolution does not follow that of other resources (especially CPU), so that researchers talked about the memory wall issue~\cite{LimCMRRW09,Shan:2018:LDD:3291168.3291175}.
Fourth, we believe that economic factors (driven by the increased competition between cloud providers, as well as rising costs of energy and hardware\footnote{We expect hardware manufacturing costs to increase due to various factors such as more complex semiconductor manufacturing processes and rarefaction of strategic raw materials. Accordingly, despite improvements in energy proportionality~\cite{barroso18datacenter}, high resource utilization will be necessary to amortize hardware acquisition costs.}, possibly increased by higher rates of environmental taxes) will push providers to resort more aggressively to resource overcommitment in production data centers in order to achieve higher resource utilization.

\textit{(Second contribution)} This paper presents \textit{Page Reference Logging} (PRL for short), a smart extension of PML for tracking both read and write page accesses in order to facilitate WSS estimation.
We show that a naive extension of PML (by simply taking into account read accesses) would lead to severe performance degradation (up to 34.8\% as indicated above).
With PRL, we allow two exclusive modes: $PRL_{PML}$ and $PRL_{PAML}$.
The former is similar to the current PML functioning, making PRL effective for live migration and checkpointing.
$PRL_{PAML}$ focuses on WSS estimation. 
In $PRL_{PAML}$ mode, read accesses are taken into account, and several loggings of the same page are also possible, making it possible to track hot pages.
Also, $PRL_{PAML}$ prevent any overhead on user VMs as follows.
VMExits related to $PRL_{PAML}$ are redirected to the privileged/controller VM's CPUs.
Notice that the privileged VM (noted pVM), which is present in almost all virtualized systems (e.g. called dom0 in Xen), belongs to the data center operator.
Thus, its utilization for hosting WSS estimation computations totally makes sense since the data center operator is the main beneficiary of this task.
Basically, $PRL_{PAML}$ logs any GPA that is at the origin of an EPT walk.
When the $PRL_{PAML}$ logging buffer is full, the actual CPU (which runs the user VM whose WSS is actually computed) sends an Inter-Processor Interrupt (IPI, a new one that we introduce) to one of the pVM's CPU, thus raising a VMExit on the latter.
By redirecting VMExits related to $PRL_{PAML}$ to the pVM, the user VM can continue its execution during the handling of that VMExits unlike in PML, thus avoiding the negative impact of user VMs.
The handler of that new IPI identifies hot pages and puts them at the disposal of a WSS estimation system.
A prototype of the latter which leverages PRL is also presented in this paper.

In summary, the paper makes the following contributions:
\begin{itemize}
	\item We present the first complete study of PML from different angles.
	We list and assess the limitations of PML in the task of WSS estimation, which is a critical issue for data center operators.
	\item In the light of our analysis, we propose \textit{Page Reference Logging} (PRL), an extension of PML which makes it effective for WSS estimation.
	Our contribution has almost the same complexity as PML and we believe it could be easily integrated by Intel.
	We prototyped PRL in Gem5 \cite{Binkert:2011:GS:2024716.2024718}, a popular hardware simulator.
	\item We also describe a WSS estimation system which leverages PRL.
	We implemented a prototype in Xen and Gem5.
	\item Using both real (HPL Linpack~\cite{hpl}, BigDataBench~\cite{BigDataBench}) and synthetic applications, we evaluated and compared our solution with VMware's WSS estimation solution (which does not rely on PRL).
	The evaluation results confirm that: (1) our solution is accurate, unlike VMware (which generates quite random values); (2) our solution does not impact user VMs, unlike VMware (which sometimes lead to VM crashing); (3) our solution is not intrusive (no guest OS code modification is required), unlike other state-of-the-art solutions~\cite{Nitu:2018:WSS:3203302.3179422,180153,Waldspurger:2002:MRM:844128.844146,Jones:2006:GMB:1168857.1168861,Lu:2007:VMM:1364385.1364388,Kim:2015:DMP:2818950.2818967}.
\end{itemize}
We make available the entire source code of PRL, PML and the WSS estimation system so that other researchers can repeat or improve our work.

The remainder of the paper is as follows.
Section~\ref{pml} introduces PML.
Section~\ref{motivations} studies PML's performance.
Section~\ref{prl} presents PRL and a WSS estimation system which leverages PRL.
Section~\ref{evaluations} presents the evaluation results.
Section~\ref{rw} presents the related work.
Section~\ref{conclusion} concludes the paper.
\section{Background}
\label{pml}

\begin{figure}[ht]
	\centering 
	\includegraphics[width=\linewidth]{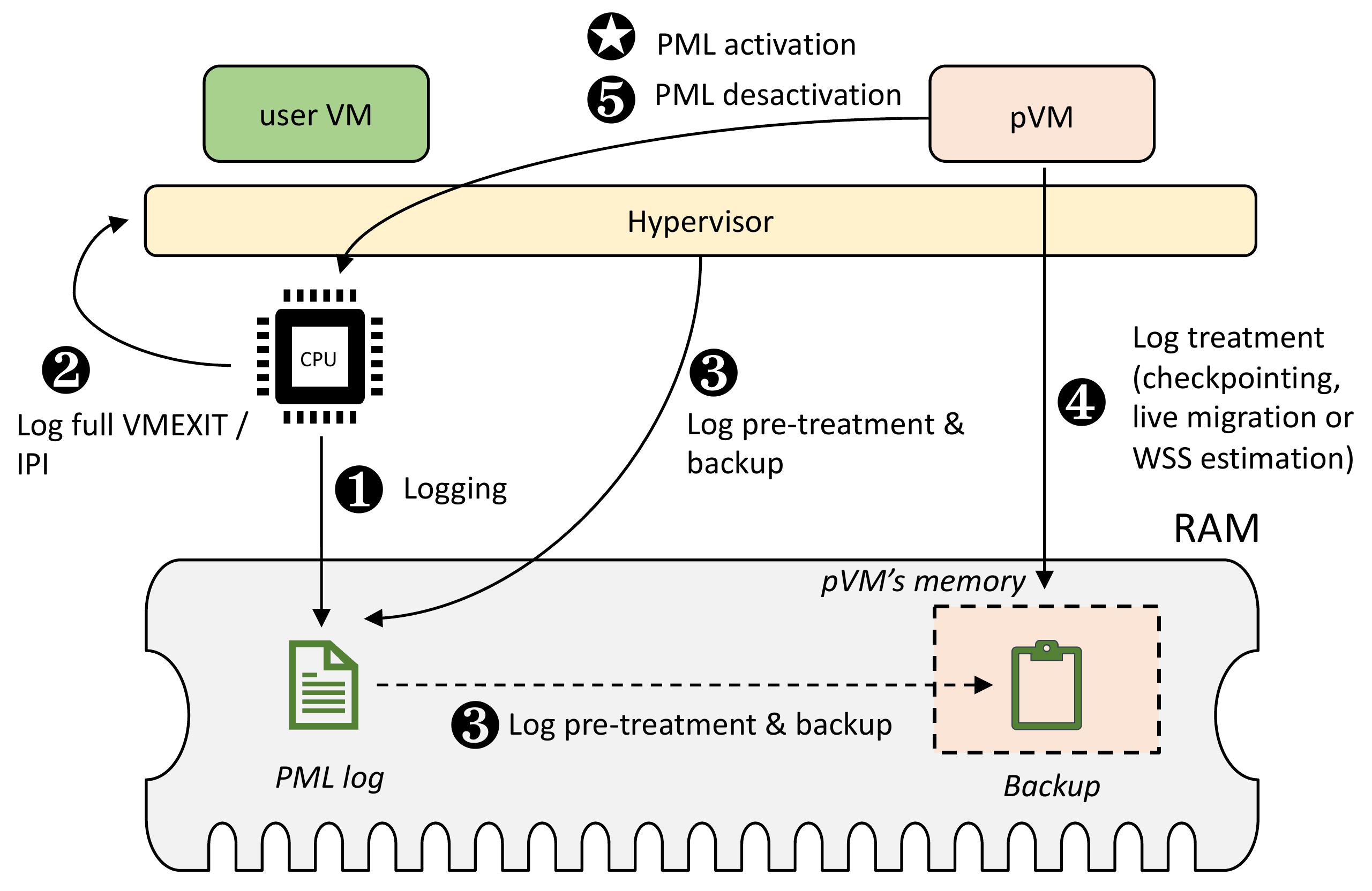}
	\caption{Basic utilization of PML for improving a virtualization operation (live migration, checkpointing and WSS estimation).}
	\label{fig:pml_actual}
\end{figure}
\subsection{Page Modification Logging (PML)}
\label{pmlItself}
PML \cite{pml} is an Intel Hardware-Assisted Virtualization (HAV) feature for memory page tracking.
It relies on Extended Page Table (EPT)~\cite{Uhlig} memory virtualization.
PML requires specific changes in the \ac{VMCS}\footnote{VMCS is a control structure associated with each vCPU when a VM runs in the Hardware-Assisted Virtualization (HAV) mode.}.
A new 64-bit VM-execution control field called \textit{PML address} is introduced.
The \textit{PML address} points to a 4KB aligned physical memory page called \textit{PML logging buffer}.
The buffer is organized in 512 64-bit entries which store logged \acp{GPA} (see below).
A new 16-bit guest-state field called \textit{PML index} is also introduced.
The \textit{PML index} is the logical index of the next entry in the logging buffer.
Because the buffer includes 512 entries, the \textit{PML index} is typically a value in the range 0-511 (starting from 511).
When PML is enabled, each write instruction which sets a dirty flag in the EPT during a page walk triggers the logging of the \ac{GPA} which is at its origin.
The \textit{PML index} is decremented after each logging operation.
Whenever the \textit{PML logging buffer} is full, the processor raises a VMExit and the hypervisor comes into play.
The logging process restarts once the \textit{PML index} is reset.
The actions taken by the hypervisor in response to that VMExit depend on the targeted goal (e.g. VM live migration).

The next section shows how PML can be integrated into the general process of several virtualization operations including VM live migration, checkpointing and WSS estimation.

\subsection{Typical PML utilization architecture}
\label{pmlUtilization}
Fig.~\ref{fig:pml_actual} shows the general functioning of a machine which relies on PML for improving a virtualization operation (e.g., live migration).
The figure shows on the one hand the user VM (green) which is the target of the virtualization operation.
The privileged VM (noted pVM)\footnote{The pVM is called \textit{dom0} in Xen~\cite{xen}, \textit{host OS} in KVM~\cite{kvm}, \textit{parent partition} in Hyper-V~\cite{hyperV}, and \textit{Service Console} in VMware~\cite{vmkernel,vmWareForum}} runs the system which implements the virtualization operation.
The execution of that system typically begins by activating PML for the target user VM (Fig.~\ref{fig:pml_actual}, \ding{74}).
Then the CPU of that VM can start logging GPAs (Fig.~\ref{fig:pml_actual}, \dingnum{1}).
On \textit{PML logging buffer} full, the CPU raises a VMExit which traps inside the hypervisor (Fig.~\ref{fig:pml_actual}, \dingnum{2}).
The handler of that VMExit does a certain task (e.g., copy the content of the \textit{PML logging buffer} to a larger buffer which is shared with the pVM (Fig.~\ref{fig:pml_actual}, \dingnum{3})).
Then the \textit{PML index} is reset to 511 and the VM resumes (VMEnter).
The system which implements the virtualization operation (in the pVM) periodically operates on the results generated by the log full handler (Fig.~\ref{fig:pml_actual}, \dingnum{4}).
This is done in respect with the virtualization operation, e.g., remigrate dirty pages in the case of live migration.
Once the virtualization operation ends, PML is disabled (Fig.~\ref{fig:pml_actual}, \dingnum{5}).

\section{PML study}
\label{motivations}
In this section we study PML from three angles: power consumption, efficiency, and performance overhead.
The two latter angles are studied under the execution of checkpointing, live migration, and WSS estimation operations.

\subsection{Experimental environment}
\label{testbed}
We realized the experiments on a machine whose characteristics are:
Single socket Intel(R) core (TM) i7-3768, 16GB memory, 500GB SSD, 4-way 64 TLB entries.
We used Xen 4.7 as the hypervisor and Linux 4.15.0 for the guest kernel.
Regarding the applications which run inside the VM, we used HPL Linpack~\cite{hpl}, BigDataBench~\cite{BigDataBench} (read, write and sort applications, 10GB data set size), and a synthetic application.
The template of the latter is presented in Fig.~\ref{fig:template}, interpreted as follows.
It consists in parsing an array several times.
Each array entry points to a 4KB (size of a memory page) data structure.
The operation type (read or write) performed on an array entry is decided according to a write intensity parameter (\texttt{wi}) which represents the proportion of write operations.
Otherwise indicated, the array uses 400MB of memory and the VM has one vCPU and 1GB of memory.

\subsection{Power consumption}
\label{energyConsumption}
\begin{figure}[ht!]
    \centering  
    \scriptsize
    \includegraphics[width=1\linewidth]{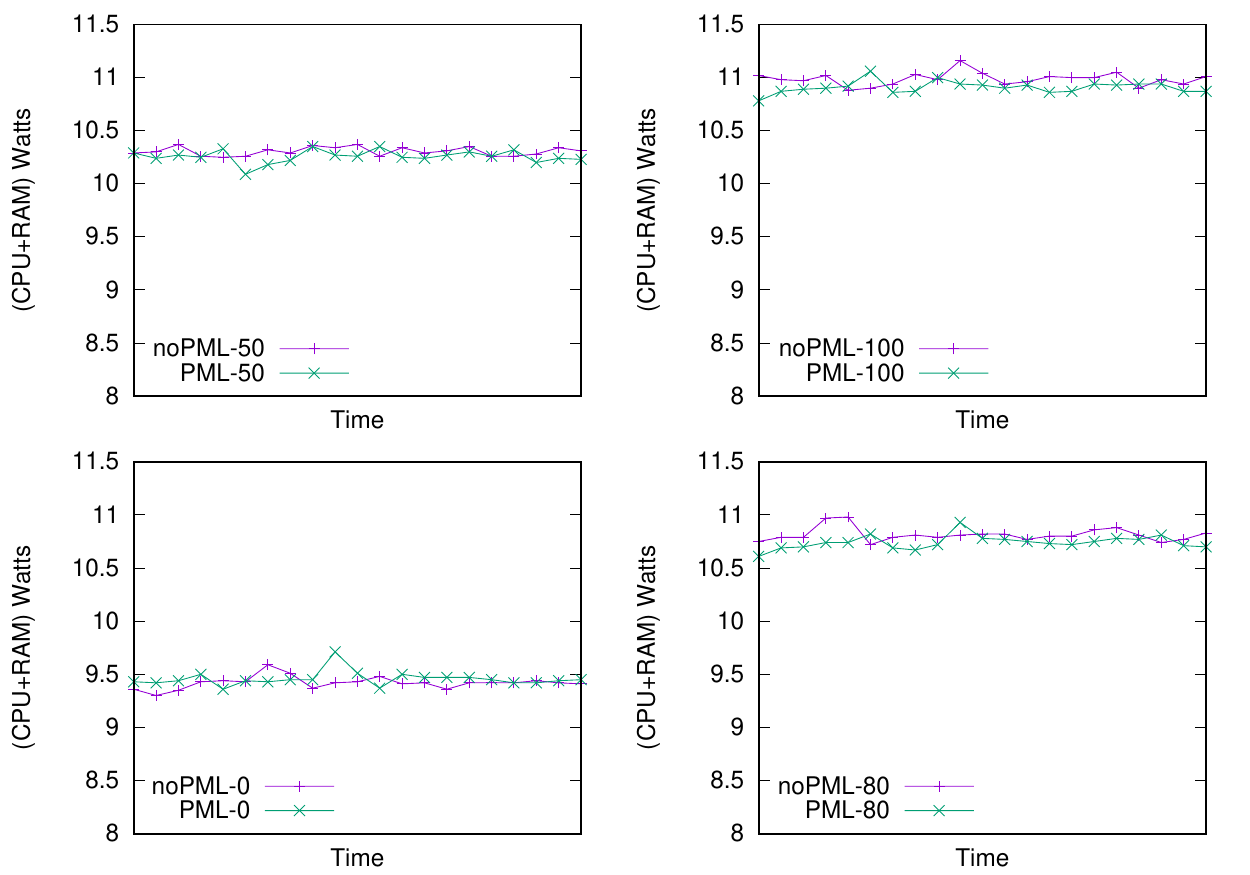} 
    \caption{Power consumption related to PML. In (no)PML-x, x represents the proportion of write operations performed by the synthetic application.
    }
	\label{fig:energyImpact}
\end{figure}

We evaluated the potential additional power consumption incurred by PML with all benchmarks.
For the synthetic application, we vary the proportion of write operations (0\%, 50\%, 80\%, and 100\%).
We use the \emph{turbostat} tool in the pVM (dom0 in Xen) to collect both CPU and memory power consumption results, presented in Fig.~\ref{fig:energyImpact}.
The latter only presents results for the synthetic application, which is very representative.
Power consumption incurred by PML is almost nil.
Note that in this specific experiment, the synthetic application is CPU and memory intensive while the VMExit handler consumes very few resources.

\begin{figure}
\small
\center
\begin{lstlisting}[language=C, 
backgroundcolor = \color{lightgray}, 
xleftmargin=.08\textwidth,
numbers=left,
  firstnumber=1,
  linewidth=8cm,
  basicstyle=\footnotesize,
  numberfirstline=true]
int app(int wi){
  unsigned long opType,nbOps=0;
  tab=malloc(..);
  for(i=0;i<D;i++){
    for(j=0;j<N;j++){
      opType=random(100);
      if(opType<wi)
        tab[j]=val;
      else
        val=tab[j];
        nbOps++;
    }
    /*opThroughput: the number of
    operation per nanosec*/
    printf("%f\n",opThroughput);
  }
  free(tab);
}
\end{lstlisting}
\caption{The synthetic application template. Its performance metric is the number of operations per nanosecond.}
\label{fig:template}
\end{figure}

\begin{figure}
    \centering  
    \scriptsize
\includegraphics[width=1\linewidth]{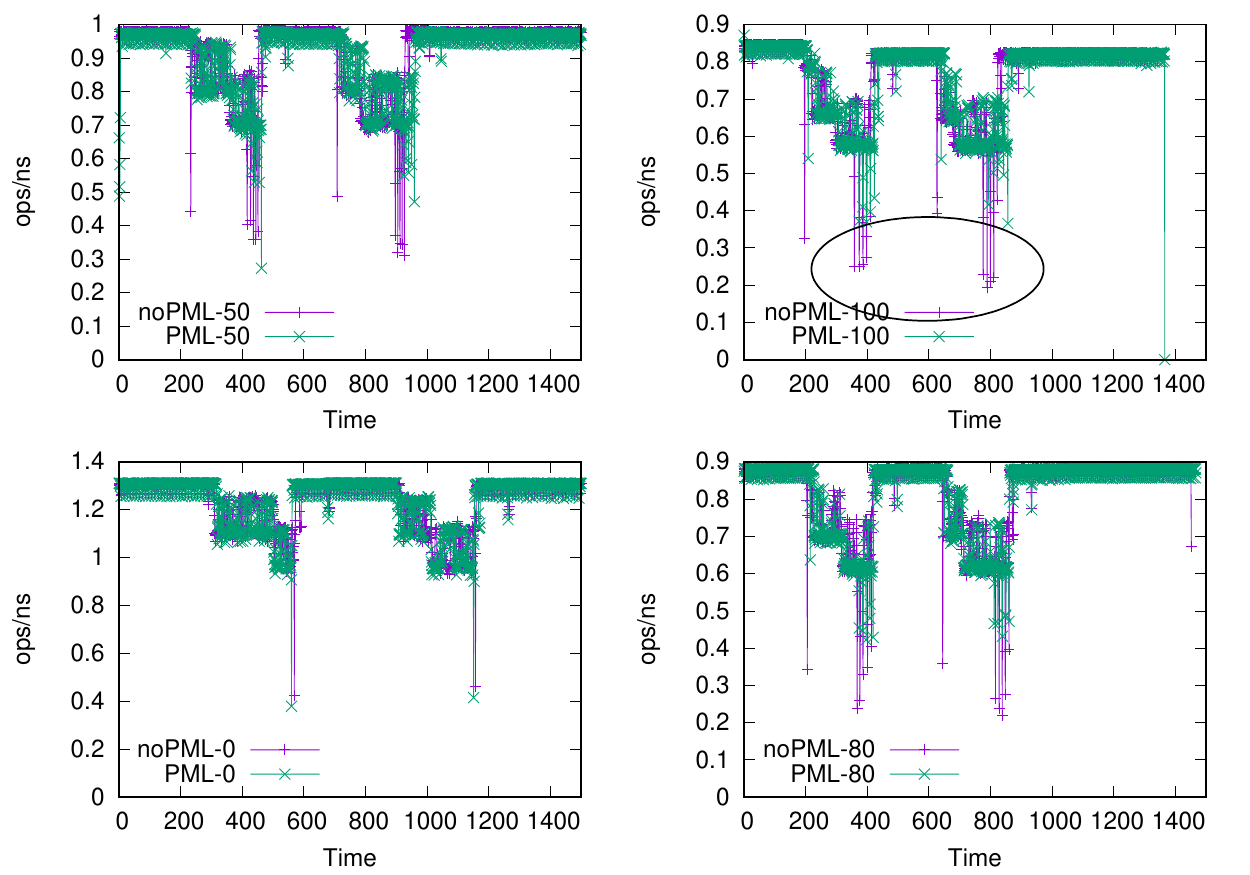}
\begin{tikzpicture}
  \begin{axis}[
                height=4cm,
                width=8cm, 
    ybar ,   
	bar width=10pt,	
	ymin=0,
	ymax=1.5,			
    legend style={at={(0.5,1.15)},
    legend cell align={left},
      anchor=north,legend columns=-1},
    ylabel={perf. improvement},
    symbolic x coords={0,50,80,100},
    xtick=data,
    ylabel near ticks,
    x tick label style={rotate=90,anchor=east},
    nodes near coords,nodes near coords align={vertical},  
every node near coord/.append style={rotate=90, anchor=west},  
    ]
    \addplot +[draw=red, pattern color=red,pattern=north east lines,thick] coordinates {(0,0.06) (50,0.140317096235487) (80,0.385758810265979) (100,0.95179209589481)};
  \end{axis} 
\end{tikzpicture} 
    \caption{PML benefits during live migration. The first four curves show the number of operations per nano second during the execution of the application while two live migrations are performed. We evaluated different write intensity values (0\%, 50\%, 80\%, and 100\%).
    The last curve (histogram) summarizes the improvement brought by PML.
    }
	\label{fig:pmlCheckpointMigration-1}
\end{figure}
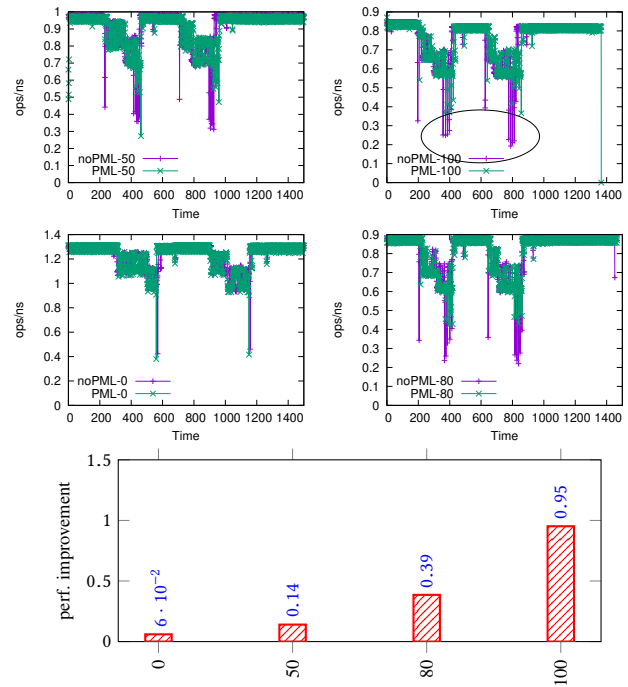

\begin{figure}
    \centering  
    \scriptsize
\begin{tikzpicture}
  \begin{axis}[
                height=4cm,
                width=8cm, 
    ybar ,   
	bar width=10pt,	
	ymin=0,
	ymax=15,			
    legend style={at={(0.5,1.15)},
    legend cell align={left},
      anchor=north,legend columns=-1},
    ylabel={\emph{save(...)} improvement},
    symbolic x coords={0,50,80,100},
    xtick=data,
    ylabel near ticks,
    x tick label style={rotate=90,anchor=east},
    nodes near coords,nodes near coords align={vertical},
every node near coord/.append style={rotate=90, anchor=west},
    ]
    \addplot +[draw=red, pattern color=red,pattern=north east lines,thick] coordinates {(0,10.1898799173928) (50,5.99414047527835) (80,2.4312151061638) (100,0.981342988983813)};
  \end{axis} 
\end{tikzpicture}

    \caption{Improvement (execution time reduction) of \texttt{static int save(...)} method, which dictates the duration of VM live migration.
    }
	\label{fig:pmlCheckpointMigration-2}
\end{figure}
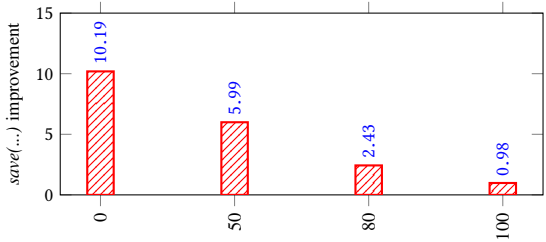

\subsection{VM migration and checkpointing}
\label{migrationCheckpointing}
VM live migration and checkpointing share some portion of code such as the one which performs memory saving (\texttt{static int save(...)} from \texttt{xc\_sr\_save.c} in Xen).
The intervention of PML in these two operations is limited to this memory saving phase.
The latter dictates the duration of the operation.
We compare the utilization of PML with the classical memory page tracking approach which consists in write protecting memory pages so that next writes lead to page faults.
We use this classical approach as the baseline for the comparison.
We are interested in two metrics:
$m_1$ the performance of the user application during checkpointing/migration and
$m_2$ the duration of the \texttt{static int save(...)} method.
$m_1$ allows to check whether PML reduces or increases the negative impact of these operations on the application while
$m_1$ tells whether PML accelerates checkpointing/migration.
To avoid noises during checkpointing, the destination file is mounted into the RAM.
We use the synthetic application for this evaluation while varying the proportion of write operations (in order to stress PML) as above.
We use this application because its behaviour is predictable in comparison with the macro-benchmark.
Two successive live migration (respectively checkpointing) operations are done during the execution of the application.

Fig.~\ref{fig:pmlCheckpointMigration-1} presents the results for $m_1$, the performance of the application (number of operations per nano second) during two consecutive live migration operations.
We can observe that even with PML, live migration still negatively impact the application performance, illustrated by the two down peaks in all curves of Fig.~\ref{fig:pmlCheckpointMigration-1}.
However, we can observe that PML slightly minimizes this impact, by 0.065\%-0.95\%.
The reduction amplitude depends on the write intensity of the workload.
The reader can see the en-cycled zone in Fig.~\ref{fig:pmlCheckpointMigration-1} as an example:
the PML-100 curve (meaning PML is enabled) is over noPML-100 curve (the classical solution).
The amelioration increases with the write intensity:
compare the read only (noPML-0 and PML-0) with the write only (noPML-100 and PML-100) results in Fig.~\ref{fig:pmlCheckpointMigration-1}.

Fig.~\ref{fig:pmlCheckpointMigration-2} focuses on $m_2$, on the execution time of \texttt{static int save(...)}, which affects the migration time.
We can see that PML reduces $m_2$ during live migration by 0.98\%-10.18\%.
Especially, read intensive applications migrate much faster when PML in used.
Notice that it is very important to accelerate live migration because it allows to quickly free a machine for maintenance, place in quarantine a corrupted VM, etc.

In contrast, we observed that PML does not actually improve checkpointing.
This is because VM checkpointing suspends the execution of the VM, which is not the case during live migration.
We think that live checkpointing would likely take advantage of PML.

\subsection{WSS estimation}
\label{wssEstimation}
Building on our experience, we implemented in Xen a WSS estimation prototype which relies on PML.
This WSS algorithm is summarized in Section~\ref{wssEstimationSystem}.
Briefly, it consists in activating PML on the VM.
Then, logged GPAs are collected and the WSS is computed as follows.
The working set of the VM is reached when no new GPAs are seen in the buffer log.
The WSS is the total number of distinct collected GPAs.

However, we concluded that it is not possible to provide an accurate WSS system using the current PML design.
We identified three limitations (noted $L_i$) of the latter. 
Two of them are hard limitations, making PML not able to accurately estimate a VM WSS.
The latter is a soft limitation which makes PML unfair for cloud users and their VMs.

\paragraph*{($L_1$) - \textit{hard}} PML does not log all accessed pages.
In fact PML only logs \acp{GPA} which are at the origin of the dirty bit setting (hence the name PML).
It means that only page modifications are recorded.
However, the WSS of a VM should include both read and write accesses.
We experimentally assessed this PML limitation using the synthetic application presented above (Fig.~\ref{fig:template}), in which we eliminate the first loop.
We observed for all executions that the estimated WSS was always the write proportion while the correct WSS is the array size.
Not counting read accesses within the WSS leads to memory under-estimation, thus performance degradation (maybe crash) for the VM.

\paragraph*{($L_2$) - \textit{hard}} Using PML, it is not possible to only track hot pages, which are the relevant ones for WSS estimation.
A page is said "hot" if it is referenced several times during a short period of time.
In respect with the current PML design, an accessed page is logged only once.
Using this design for WSS estimation, it is not possible to distinguish "hot" and "cold" pages, resulting in the over-estimation of the actual VM memory needs.
To assess this limitation, we added to the synthetic application a for loop at the beginning which modifies all the array entries.
The remaining application code works on a small portion of the array (noted M, $M<N$).
In this case, M is the correct WSS.
Using PML, N is reported as the WSS, leading to memory waste.\\
This current PML design is sufficient for live migration and checkpointing because these two operations are only interested in tracking modified pages.
(Counting the number of modifications for a given page is not necessary).

\paragraph*{($L_3$) - \textit{soft}} The handling of PML logging buffer full events should not be done by the CPU of the VM whose WSS is estimated.
In fact, depriving the user VM from its CPU quota is unfair because WSS estimation is only beneficial for the data center operator.
The reader could legitimately say that this limitation is also true for checkpointing and migration.
First checkpointing is done while the VM is suspended.
Second, it has been proven~\cite{Nitu:2017:SBQ:3050748.3050758} that a slight reduction of the CPU time used by the migrated VM accelerates live migration.

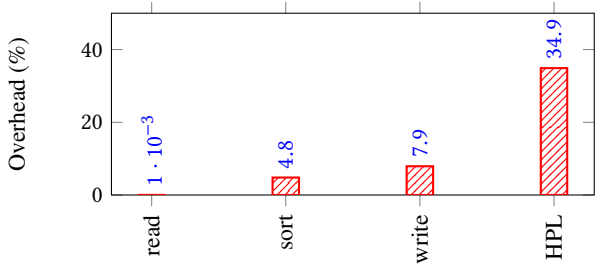
\begin{figure}[ht!]
    \centering  
    \small 
\begin{tikzpicture}
  \begin{axis}[
                height=4cm,
                width=8cm, 
    ybar ,   
	bar width=10pt,	
	ymin=0,
	ymax=50,	
	ymax=50,		
    legend style={at={(0.5,1.15)},
    legend cell align={left},
      anchor=north,legend columns=-1},
    ylabel={Overhead (\%)},
    symbolic x coords={read,sort,write,HPL},
    xtick=data,
    x tick label style={rotate=90,anchor=east},
    nodes near coords,nodes near coords align={vertical},
every node near coord/.append style={rotate=90, anchor=west},    
    ]
    \addplot +[draw=red, pattern color=red,pattern=north east lines,thick] coordinates {(read,0.001) (sort,4.8) (write,7.9) (HPL,34.9)};				
  \end{axis}   
\end{tikzpicture}
    \caption{The impact of PML when it is used for WSS estimation. %
    For BigDataBench applications, the dataset is 10GB.
    We can notice with up to 34.8\% of performance degradation for HPL Linpack}
	\label{fig:pmlImpact}
\end{figure}

We measured the overhead of the current PML design to estimate the WSS of applications from BigDataBench~\cite{BigDataBench} (read, write and sort) and HPL Linpack~\cite{hpl} (detailed in Section~\ref{evaluations}).
To this end, we run each application with and without PML and compute the overhead, presented in.
Fig.~\ref{fig:pmlImpact}.
We can see that read intensive applications are not impacted by PML because it does not solicit PML.
However, other workloads are impacted by PML, with up to 34.8\% for HPL Linpack.

\subsection{Synthesis}
\label{synthesis}
The main conclusion of this study is as follows.
First, PML incurs almost no power consumption overhead.
Second, PML reduces VM live migration time.
Third, live checkpointing could take benefit from PML.
Fourth, PML slightly reduces the performance overhead on applications during live migration.
Fifth, PML does not allow accurate WSS estimation.

Working set estimation is useful for a variety of tasks in the data center, including:
\begin{itemize}
	\item \textit{fast VM restore, after checkpointing}.
	As shown by~\cite{Zhang:2011:FRC:1952682.1952695}, VM restore can be accelerated if only the working set of the VM is restored, instead of the entire VM memory.
	\item \textit{optimal processor cache partitioning}.
	Intel Cache Allocation Technology is a hardware feature which allows to partition and isolate the last level processor cache.
	The determination of the cache size for each VM is a very tricky task.
	\cite{intelCatRealTime} showed that the WSS of a VM is the optimal value for its cache size.
	\item \textit{optimal resource utilization using memory overcommitment}.
	Resource waste is one of the main challenges in todays data centers.
	Memory overcommitment~\cite{Overcommit}, which is the capability to dynamically adjust each VM memory size according to its actual needs, has been demonstrated by several research work~\cite{Nitu:2018:WZP:3190508.3190537} to be a very promising approach for reducing resource waste.
\end{itemize}
The next section presents an extension of PML which makes it effective for WSS estimation.
\section{Page Reference Logging}
\label{prl}
This section presents \textit{Page Reference Logging} (PRL for short), an extension of PML for making the latter usable for WSS estimation.
Conceptually, PRL includes two innovations:
\begin{itemize}
	\item the capability to track both read and write accesses;
	\item the redirection of log full events to pVM's CPUs, instead of user VMs.
\end{itemize}

\begin{figure}[ht]
	\centering
	\includegraphics[width=1.4\linewidth]{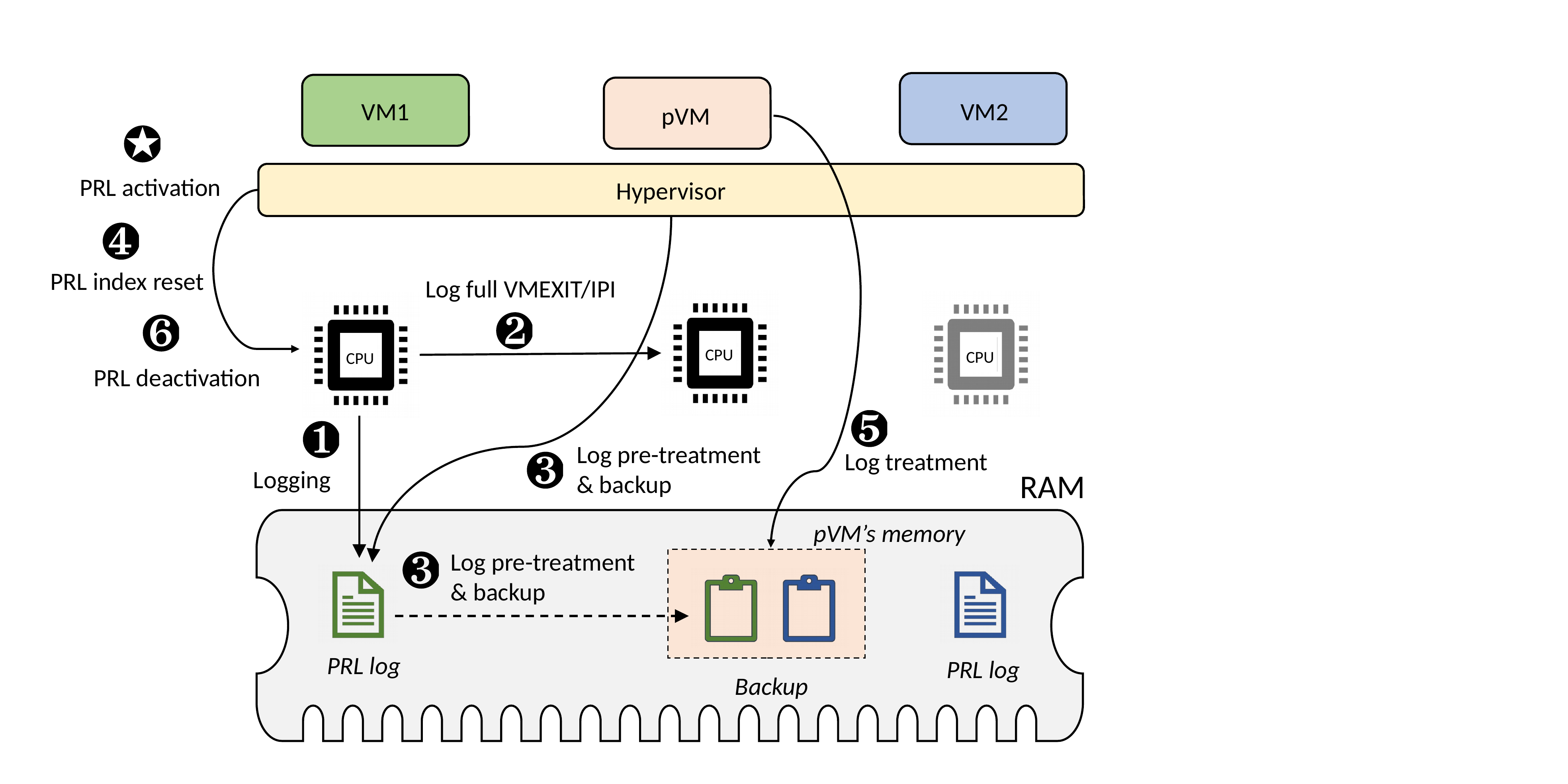}
	\caption{PRL, the design that we propose. It works in two exclusive modes: $PRL_{PML}$ and $PRL_{PAML}$. The latter mode is only presented in this figure.
	It allows accurate WSS estimation without impacting user VMs.}
	\label{fig:prlDesign}
\end{figure}
\subsection{Design}
\label{apalDesign}
PRL allows the processor to function in two exclusive modes: $PRL_{PML}$ and $PRL_{PAML}$ (PAML stands for Page Access and Modification Logging).
In $PRL_{PML}$, the processor works in the same way as the current PML design, thus allowing PRL to satisfy live migration and checkpointing requirements.
Concerning $PRL_{PAML}$, it is enabled in the same way as $PRL_{PML}$ with the sole difference that the system software should set a new bit of the \textit{Secondary Processor-Based VM-Execution Controls} (e.g. bit 26).
This section focuses on the description of $PRL_{PAML}$ (see Fig.~\ref{fig:prlDesign}), which tackles all requirements related to WSS estimation.

First, a new 16-bit host-state field called "\textit{log full handler CPU}" indicates the index of the CPU to which an interrupt is sent when the PRL log buffer is full.
A new 8-bit host-state field called "log full vector" indicates the interrupt vector which will be executed by the target CPU on log full interrupt reception.
This destination CPU should belong to the pVM so that the latter services as the execution room of the WSS estimation system (Fig.~\ref{fig:prlDesign}, \dingnum{2} - \dingnum{5}), presented in Section~\ref{wssEstimationSystem}.
By this way, $PRL_{PAML}$ avoids to schedule out the VM whose WSS is being estimated.
Remember that the pVM belongs to the data center operator, thus its utilization for WSS estimation makes sense.

Second, for every GPA which is input of the PRL process (which starts after an EPT walk, on TLB miss), the following algorithm takes place:
\begin{enumerate}
	\item If all bits of \textit{PRL index} are zero, it means that the PRL logging buffer has been detected to be full;
	Then \textit{PRL index} is decremented and an interrupt is sent to the pVM's CPU which is responsible for handling log full events.
	The PRL process ends without the interruption of the VM whose WSS is being estimated.
	The processor starts logging again upon the reset of \textit{PRL index} by the system software, especially the log full event handler (see Section~\ref{logFullEventRedirect}).
	\item If the \textit{PRL index} is negative, it means that the log full handler is executing. So we don't log the \texttt{gpa} and the PRL process ends;
	This means that PRL misses some \texttt{gpas} during the treatment of the execution of the handler.
	We claim that this does not affect the estimation of the WSS because if a missed \texttt{gpa} belongs to the working set, it is likely to be seen in the near future (after the re-enabling of the PRL mechanism) since it is hot.
	Otherwise, the \texttt{gpa} is cold and its loss is not an issue.
	The evaluation results confirm our claim.
	\item Otherwise, \textit{PRL index} is decremented and \texttt{gpa} is logged.
	This is done regardless the value of the dirty flag.
	By this way, $PRL_{PAML}$ can log both accessed and modified pages.
	In addition, $PRL_{PAML}$ can log the same page access several times.
\end{enumerate}

\subsection{Full event redirection to pVM's CPUs}
\label{logFullEventRedirect}
To this end, the actual processor sends (through its LAPIC) an IPI (Inter-Processor Interrupt) to the pVM's CPU which has been designated at VMCS configuration time.
The LAPIC is configured with the identifier of that target CPU.
We have introduced a new interrupt vector which points to the full event handler.
This mechanism is similar to "\textit{Lightweight inter-core notifications}" introduced by Jefrey C. Mogul et al. in \cite{Mogul}.

Since the full event handler should run in the VMX root mode because it deals with VMCS data structures, the target processor must trigger a VMExit upon receiving the IPI.
This behavior is enforced by setting bit 0 of the \textit{Pin-Based VM-Execution Controls} of all pVM's vCPUs.
Using this configuration, any external interrupt sent to any CPU which runs pVM  trigger a VMExit.

\subsection{Full event handling on pVM's CPUs}
\label{logFullEventHandling}
The basic actions that the handler of the log full event should implement are as follows.
First of all, it masks the interrupt related to the log full event.
It then transfers the content of all PRL logging buffers which are full (for all VMs) to larger buffers.
Notice that each VM is assigned a dedicated large buffer, which is unique even for a multi-vCPU VM.
During this transfer, the handler accumulates for each GPA the number of times it has been seen in the PRL logging buffer.
This way, the WSS estimation system (see the next section) could determine hot pages.
After the transfer of PRL logging buffers, the handler should reset (to the initial value) the \textit{PRL index} of all VMCSs which were detected as full.
Notice that the modification of a \textit{PRL index} only concerns its VMCS memory region, not the corresponding processor internal register.
In fact, the synchronization of the VMCS memory region and the processor registers is not automatic.
To enforce this, we introduce a new instruction which allows to refresh the internal VMCS state of a specific processor using its corresponding VMCS memory region.
The execution of the handler should end by the unmasking of the full interrupt.\\
Notice that this algorithm allows to handle several log full events using only one generated interrupt.
The algorithm is inspired by the New API (NAPI) implemented in modern Linux kernels for handling network packet reception~\cite{napi}.

\subsection{A PRL-based WSS estimation system}
\label{wssEstimationSystem}
Memory overcommitment~\cite{Waldspurger:2002:MRM:844128.844146} consists in dynamically adjusting the amount of memory allocated to a VM according to its actual needs, thus avoiding waste.
This on-demand memory management strategy requires the data center operator to permanently estimate for each VM its current WSS (noted $M$).
The latter can then service as input of several optimization tasks in the data center, see the previous section.
This section presents a \textit{WSS estimation system} which leverages PRL.
This system runs inside the pVM.

A PRL-based \textit{WSS estimation system} launches as many WSS estimation processes as the number of tenant VMs.
Each process calculates $M$ using this equation
\begin{equation}
	M=wss \times pageSize +\varepsilon
	\label{eq:eqM}
\end{equation}
where $wss$ is the number of hot pages, $pageSize$ is the size of a memory page, and $\varepsilon$ is the size of the guest kernel footprint.

\paragraph{$wss$ computation}
The value of $wss$ is computed using GPAs logged by PRL.
The computation algorithm takes three parameters as input:
\begin{enumerate}
	\item[$\tau$:] a page which GPA has been logged at least $\tau$ times is considered as a hot page;
	\item[$\omega$:] the WSS stability duration, used to determine whether the VM has already covered its working set;
	\item[$\mu$:] the observation interval.
\end{enumerate}
The values of these parameters are decided by the external entity which launches the WSS estimation system.
\cite{Zhao:2011:LCW:2002181.2002198} presents a set of methods which can be used to determine the values of these parameters.

\begin{figure}[t!]
	\centering
	\includegraphics[width=.7\linewidth]{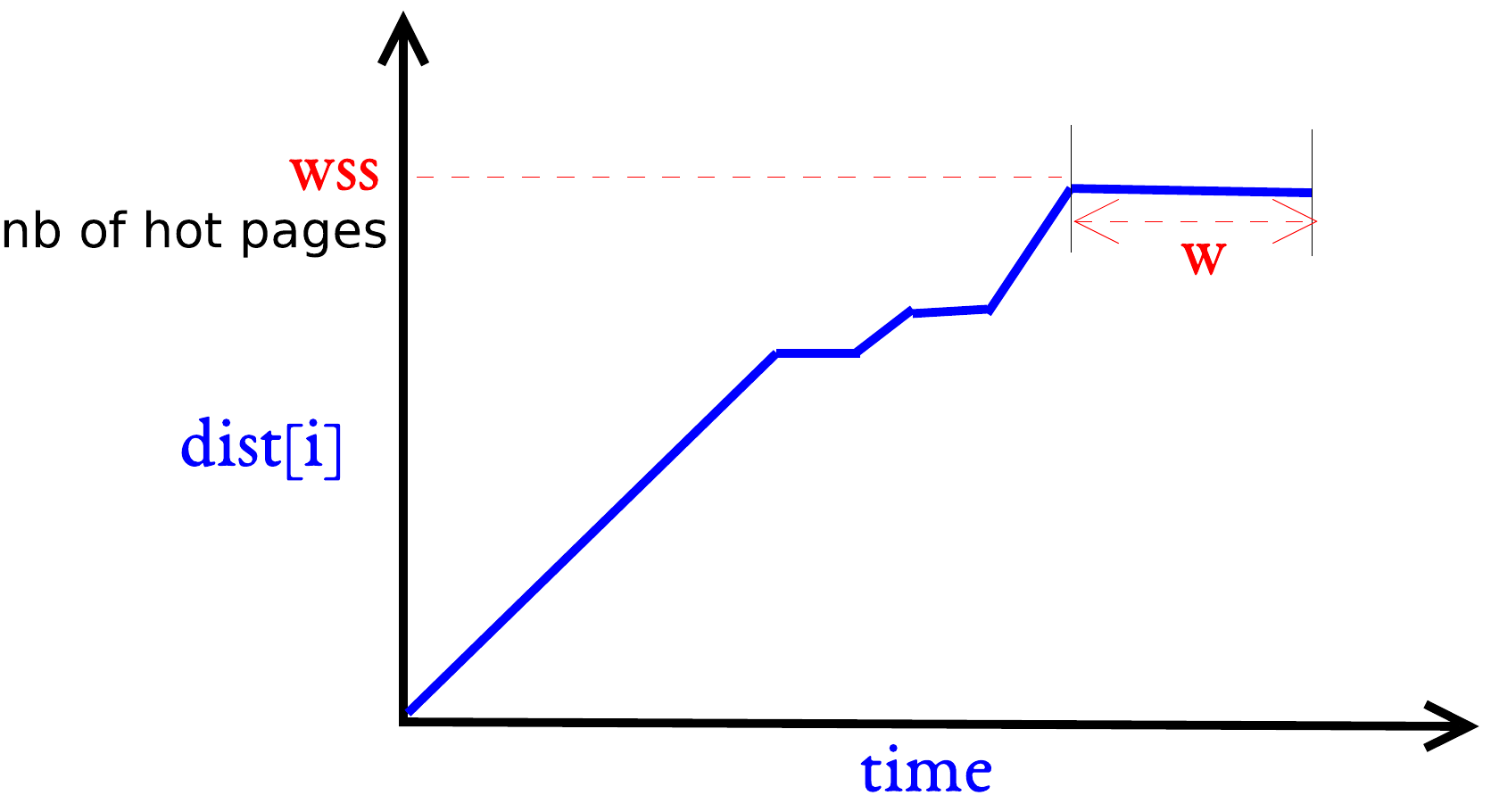}
	\caption{Illustration of the WSS estimation algorithm.
	}
	\label{fig:illustrationWSS}
\end{figure}

The estimation algorithm works as follows.
Let us note \emph{@buff} the address of the cumulative PRL buffer of the VM whose WSS is calculated.
The WSS estimation algorithm consists of a while loop.
For each iteration $i$, the number of distinct GPAs present in \emph{@buff} which have been logged more than $\tau$ times is computed and stored in $dist[i]$.
The loop ends when $dist[i]-dist[i-\omega]=0$.
This condition is true when the VM has touched/referenced all memory pages belonging to its current working set.
Otherwise, the process sleeps during $\mu$ seconds before continuing the iteration.
Fig.~\ref{fig:illustrationWSS} illustrates this algorithm.
We can see the evolution of $dist[i]$ over the time that corresponds to an increasing monotonic function.

\paragraph{$\varepsilon$ computation}
The value of $\varepsilon$ depends on the guest kernel binary.
It is estimated once by the data center operator for each kernel binary using the following algorithm:
\begin{enumerate}
	\item Starts a 2GB VM from the kernel binary;
	\item Initialize $\varepsilon$ and $curMem$ (an auxiliary variable) to 2GB;
	\item Set $curMem$ to $95\% \times \varepsilon$;
	\item Change the VM memory size to $curMem$;
	\item If the VM crashes then stop the algorithm and return $\varepsilon$;
	\item Else set $\varepsilon$ to $curMem$ and go to step 3.
\end{enumerate}
We provide a tool that automates these steps for a machine virtualized with Xen hypervisor.
\begin{figure}
\small
\begin{lstlisting}[language=C, 
backgroundcolor = \color{lightgray}, 
xleftmargin=.08\textwidth,
numbers=left,
  firstnumber=1,
  linewidth=8cm,
  basicstyle=\footnotesize,
  numberfirstline=true]
int app(){
  for(i=0;i<D;i++)
    for(j=0;j<N;j++)
      operation(tab[j]);
}
\end{lstlisting}
\caption{The second synthetic application template.}
\label{fig:listing1}
\end{figure}

\begin{figure*}[ht!]
    \centering  
    \scriptsize
\includegraphics[width=1\linewidth]{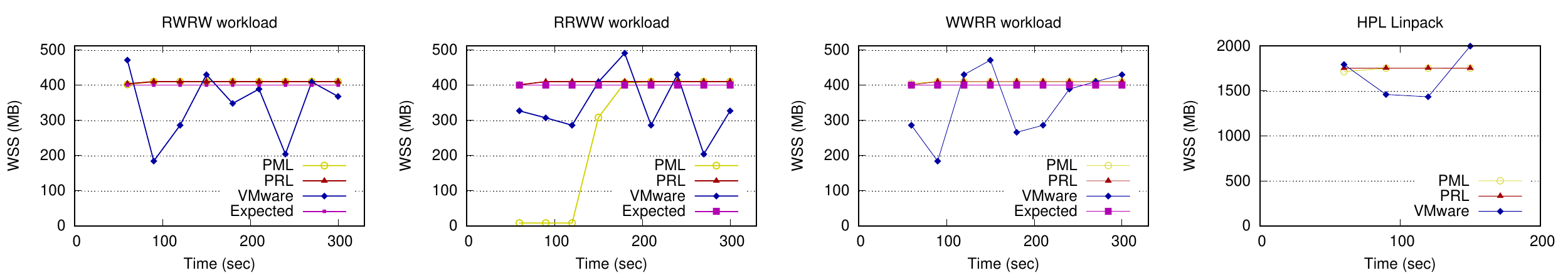} 
    \caption{Accuracy of PRL compared with PML and VMware.}
	\label{fig:accuracy}
\end{figure*}

\section{PRL Evaluation}
\label{evaluations}
This section presents the evaluation results of PRL and the WSS estimation system which relies on it.
The evaluations cover two aspects: accuracy and overhead.
The former is the capability of PRL to accurately estimate the WSS of a VM.
The WSS estimation algorithm is configured as follows: $\tau=50$, $\mu=30s$, and $\omega=120s$ (soit $4*\mu)$.
The value of these parameters were found empirically.
Concerning the overhead, we evaluated:
(1) the additional power consumption incurred by PRL,
(2) the impact incurred by PRL on the VM whose WSS is estimated (referred to as "the target VM"),
and
(3) the amount of resources consumed by the WSS estimation system inside the pVM.

\subsection{Experimental environment}
\label{newExpeEnv}
The experimental environment is the same as the one presented in Section~\ref{testbed} (which uses a real hardware), completed with the hardware simulator Gem5 \cite{Binkert:2011:GS:2024716.2024718}.
The combination of these two environments allows us to emulate a machine which implements PRL, see below.
We chose Gem5 because it is a very popular hardware simulator, which has been used by up to 97 research papers at the time this paper is written (according to \cite{gem5Utilisation}).
Gem5 allows the execution of a full Linux distribution.
We improved it in order to simulate a virtualized system.
This improvement consists in: (1) the addition of the Extended Page Table (EPT), (2) the extension of the hardware page table walker so that it performs a 2D page walk through the EPT and (3) the implementation of PRL/PML logging mechanisms.
The emulation methodology is as follows.
The application is first executed under Gem5 and the logged GPAs are collected, including logging instants given by the simulator.
The collected traces are then replayed inside the dom0 of a real virtualized environment which runs the WSS estimation system.
A dom0's CPU (noted $CPU_0$) is dedicated to the latter.
The other dom0's CPUs run processes which replay the traces, thus mimicking the functioning of a real PRL equipped machine.
Each process which replays the traces send an IPI to $CPU_0$ every time it has played N traces, N being the size of the logging buffer.
This emulation approximates the real functioning of a PRL-capable machine, depicted in Fig.~\ref{fig:prlDesign}.
In order to be fair, we also evaluated the accuracy of PML and VMware's solution following the same methodology.
Recall that VMware's WSS estimation solution consists in periodically (every 30s) selecting a sample of 100 memory pages whose present bits are invalidated.
The proportion of pages, among these selected 100 pages, which will lead to a page fault represents the WSS of the VM.

Concerning the application which runs inside the VMs, we also use a variant of the synthetic application presented in Section~\ref{testbed}.
This second variant follows the template shown in Fig.~\ref{fig:listing1}.
N allows to control the WSS (N $\times$ 4KB),
D represents the duration of the application, and
\emph{operation(tab[j])} is the operation performed on the array entry.
We consider three workload types, in respect with \emph{operation(tab[j])}:
(\emph{RWRW}) every read is followed by a write;
(\emph{RRWW}) a set of reads followed by a set of writes;
and (\emph{WWRR}) a set of writes followed by a set of reads.

\begin{figure*}[ht!]
    \centering  
    \scriptsize
\includegraphics[width=1\linewidth]{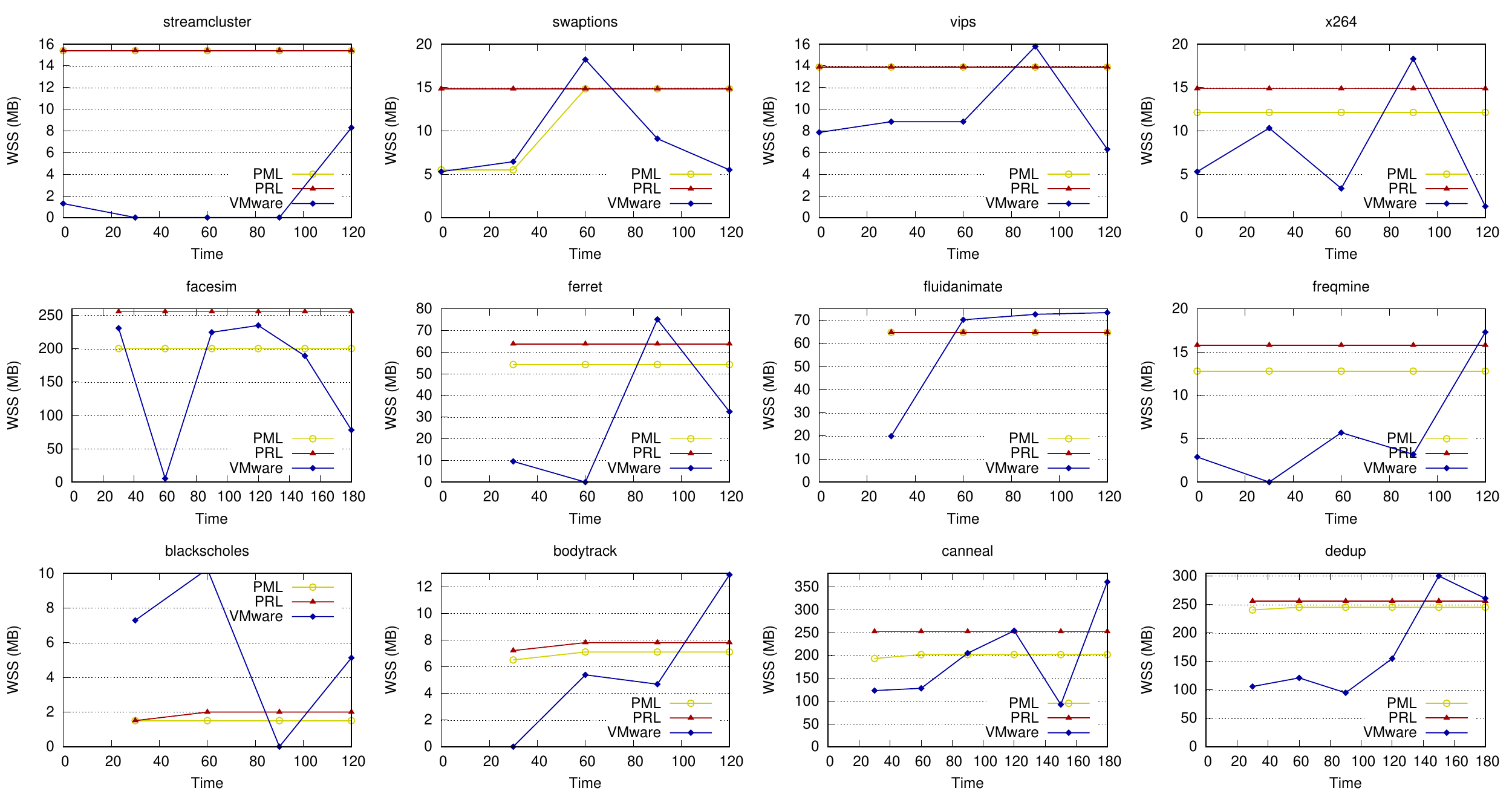} 
    \caption{WSS estimation of PARSEC applications using PRL.}
	\label{fig:accuracyParsec}
\end{figure*}

\subsection{Accuracy}
\label{accuracy}
Given a constant workload which runs inside the VM, we are interested in verifying if PRL is able to accurately estimate the WSS (the number of hot pages) of that workload.
To realize this evaluation, we used the second variant of the synthetic application and HPL Linpack.
We did not use BigDataBench applications because their execution under Gem5 never completed because the latter is a very slow simulator, thus not suitable for such bigdata applications.

Fig.~\ref{fig:accuracy} presents the results, interpreted as follows.
Concerning the synthetic application, its WSS is known in advance (400 MB). We can see that PRL is accurate for all workload types with an error margin lower than 1MB.
VMware's solution appears inaccurate, which is inline with previous research observations~\cite{Nitu:2018:WZP:3190508.3190537}.
The accuracy of PML depends on the amount of write operations.
It accurately estimates RWRW and WWRR WSS because during their execution, all the array entries are referenced by write operations.
This is not true during the first step of RRWW, explaining PML inaccuracy.

About HPL, we observed that both PRL and PML are accurate, not VMware.
We validated the accuracy (as the WSS is not known in advance) with the following protocol.
Relying on the results obtained with PRL, we dynamically adjust at runtime the VM memory size.
We observed no VM crashing and no performance degradation, meaning that either PRL has overestimated the WSS or it is accurate.
We repeated the experiment while subtracting 100MB from PRL generated values (1700 MB), which led to VM crashing.
As a conclusion, PRL obtained values were accurate.

During these experiments, we observed that the number of missed GPAs during the handling of full buffer events is negligible.
Also, a GPA which is missed at round $i$ is seen in round $i+x$ when that GPA is part of the working set.
For illustration, table~\ref{tab:missedGPAs} shows the number of missed GPAs during the execution of a synthetic application and HPL. We also observed (with two buffer sizes in table~\ref{tab:missedGPAs}) that the size of the logging buffer does not impact this result, even if the buffer processing is longer.
\begin{table*}[]
    \centering
\begin{tabular}{ |c||c|c|c|c|c|c|  }
 \hline
 Buffer size (MB) & \multicolumn{3}{|c|}{RRWW} & \multicolumn{3}{|c|}{HPL} \\
 \hline
  & \# full events & \# missed GPAs & \% of missed GPAs & \# full events & \# missed GPAs & \% of missed GPAs\\
\hline 
 512 & 17094 & 0 & 0 &  20701 & 741 & 0.04\\
 \hline
 1024 & 8543 & 213 & 0.02 & 10510 & 116 & 0.01\\
 \hline
\end{tabular}
    \caption{Number of missed GPAs during the treatment of buffer full events. We can see that the number of missed GPAs is negligible, thus not impacting the WSS estimation algorithm.}
    \label{tab:missedGPAs}
\end{table*}

After demonstrating the accuracy of PRL, we used it for estimating the WSS of PARSEC applications~\cite{parsec}.
The latter are often used by researchers, thus knowing their WSS is likely to interest several researchers.
Fig.~\ref{fig:accuracyParsec} presents the results.
We can see that PML is accurate for six applications (streamcluster, vips, fluidaminate, blacksholes, bodytrack, dedup).
VMware's WSS estimation solution leads to inaccurate values.
By making the implementation of PRL under Gem5 publicly available, researchers can use it for estimating the WSS of other benchmarks.

\subsection{Overhead}
\label{overhead}

\subsubsection{Overhead on the VM which WSS is computed}
This overhead is the total number of CPU cycles used by PRL circuitry (noted $T^{Circuitry}_{PRL}$).
Having not yet a PRL machine, we assume that $T^{Circuitry}_{PRL}$ equals $T^{Circuitry}_{PML}$, according to the slight difference between the two modes.
Therefore, to evaluate the overhead of PRL, we used a PML-capable machine on which the VM runs a read only application. This scenario avoids VMExits related to PML, thus reproducing the effect of PRL on the VM (since VMExits in PRL are executed on a different core).
We repeated this experiment while varying the number of tenant VMs in order to increase the pressure on the TLB.
Recall that PML (as well as PRL) takes place during page table walk, on each TLB miss.
We are interested in the performance difference of the application with and without PML.
We measured no performance difference (about 0.001\%), meaning that PRL will likely incur no performance degradation on the VM which WSS is computed.

\subsubsection{Resource utilization in the pVM}
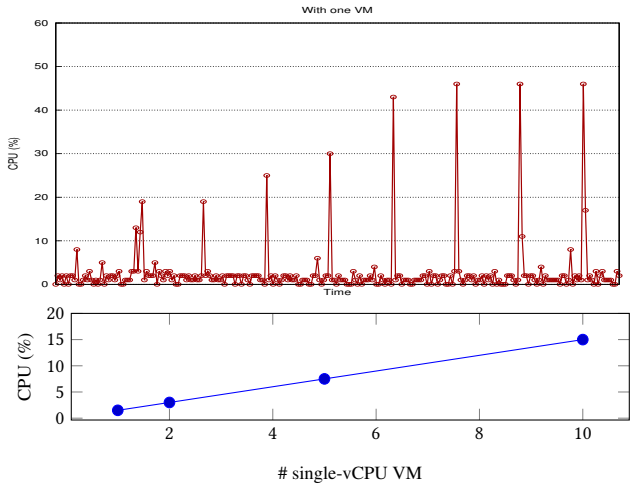
\begin{figure}[ht!]
    \centering  
    \small 
\includegraphics[width=8.5cm, height=4cm]{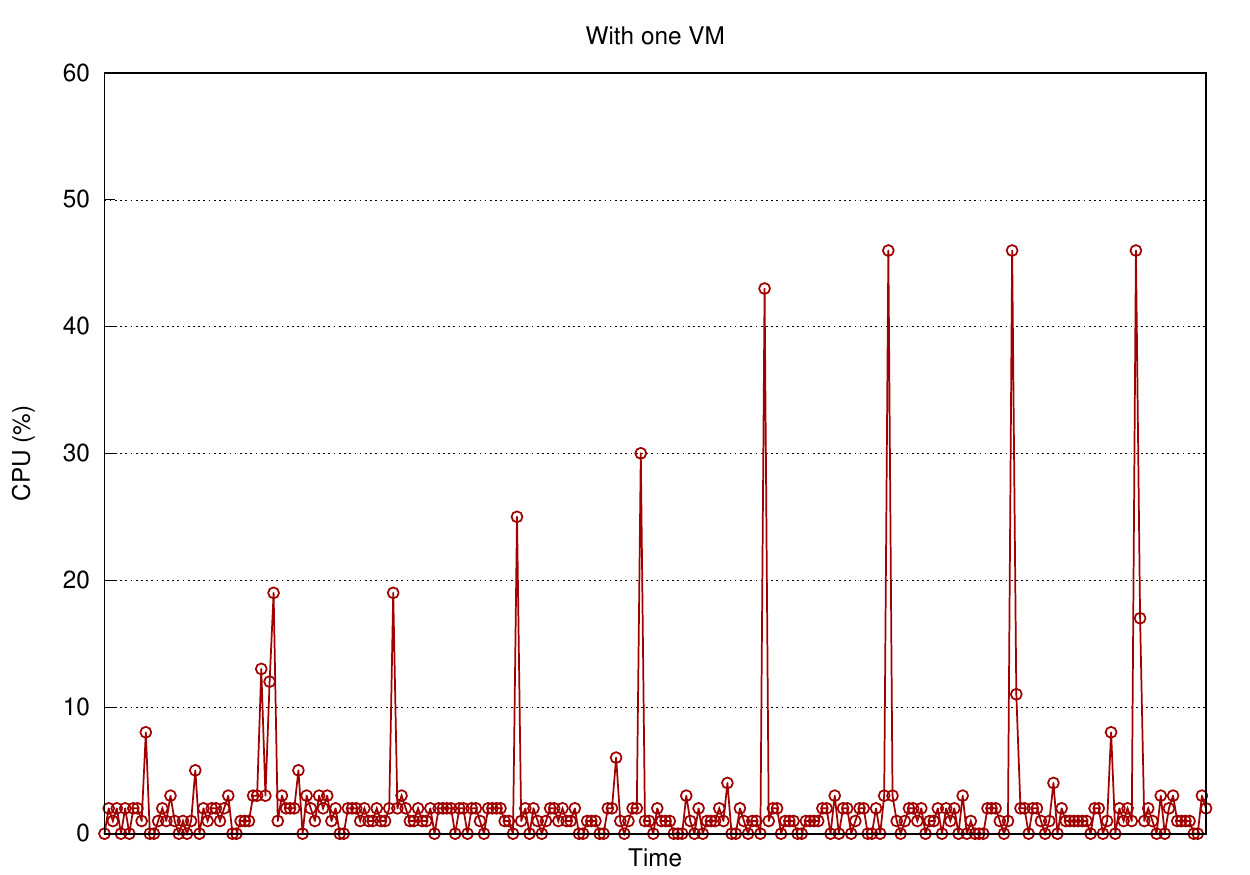}
\begin{tikzpicture}\scriptsize
  \begin{axis}[
                  height=3cm,
                width=9cm,
                legend style={at={(0.8,1.3)}},
    ymax=20,  
    ylabel near ticks,
    legend cell align=left, 
    ylabel=CPU (\%),
    xlabel=\# single-vCPU VM,
    ]
    \addplot coordinates {
        (1,1.5) (2,3) (5,7.5) (10,15)
    };
  \end{axis}
\end{tikzpicture}

    \caption{(top) Percentage of CPU time consumed by the WSS estimation process inside the pVM during the execution of a single-vCPU VM. (bottom) The average consumption when varying the number of VMs.
    }
	\label{fig:cpuConsumption}
\end{figure}

We focus on CPU consumption.
To this end, we used the emulated environment.
The WSS estimation system is dedicated a CPU core.
The write intensive synthetic application is used for this experiment because it is the worse case.
We performed the experiment with one VM.
Fig.~\ref{fig:cpuConsumption} top presents the percentage of CPU consumed by the WSS estimation process in the dom0 for a single-vCPU VM.
Only a representative portion of results is presented.
The CPU consumed during the treatment of the buffer full event (computation of $dist[i]$, see Section~\ref{wssEstimationSystem}) increases (with the size of the cumulative buffer) until the WSS is discovered.
However, we can observe that the CPU is most of the time idle, waiting for the PRL logging buffer full event.
In average, the total CPU time consumed by the WSS estimation process for a single-vCPU VM is less than 1.5\%.
We repeated the same experiment while varying the number of single-vCPU VMs.
Fig.~\ref{fig:cpuConsumption} bottom shows that the average percentage of CPU consumed by the WSS estimation system increases linearly.

\subsubsection{Power consumption}
Power consumption overhead due to PRL corresponds to the power consumed by $T^{Circuitry}_{PRL}$.
It is equal to the power consumed by $T^{Circuitry}_{PML}$.
Therefore, we evaluate $T^{Circuitry}_{PRL}$ using a PML-capable machine on which a read intensive application runs.
This experiment is the same presented in Section~\ref{energyConsumption}, see noPML-0 and PML-0 curves in Fig.~\ref{fig:energyImpact}.
Recall that we observed almost no additional power consumption due to PML, thus PRL.

\section{Related work}
\label{rw}
This section presents the related work.

\paragraph{Hardware assisted virtualization (HAV)}
Contrary to A. Baumann's conclusion in his HotOS'17 paper~\cite{Baumann:2017:HNS:3102980.3103002}, we believe that virtualization is the System's sub-domain which mainly influences hardware architecture research.
In fact, HAV contributions have evolved at the rhythm of the limitations of software solutions.
Notably, Extended/Nested Page Table~\cite{Bhargava} has been introduced for addressing the tremendous number of context switches caused by shadow paging~\cite{Waldspurger:2002:MRM:844128.844146}.
In summary, several hardware features (e.g., Intel VT~\cite{intelVT}, AMD VT~\cite{amdVT}, AMD SEV~\cite{amdSEV}, VMFUNC~\cite{vmfunc}, Intel CAT~\cite{intelCAT}, APICv~\cite{apicv}) have been integrated inside CPU, memory subsystems, I/O devices and many other motherboard components by hardware manufacturers these recent years for achieving basic virtualization functionalities.
An extreme application of the HAV approach has been proposed by E. Keller with NoHype~\cite{Keller:2010:NVC:1816038.1816010} which is a hardware only hypervisor.
In 2017, Amazon anounced its new hypervisor called Nitro~\cite{liguori18powering}, which can be seen as a concretization of the NoHype vision.
In the academia, a lot of efforts have been made in the topic of memory virtualization~\cite{Barr,Yaniv,Ahn,Wang2011,doItYourSelf,Gandhi,Gandhi2014,lim17neve} to minimize the overhead of the 2D page walk imposed by EPT.

\paragraph{Page tracking}
The most popular approach for page tracking consists in denying access to memory pages which need to be monitored so that next accesses trap inside the system software (hypervisor or OS).
This approach is used by the majority of checkpointing, live migration and WSS estimation solutions.
Very few research works have investigated hardware features for page access tracking.

Pin Zhou et al.~\cite{Zhou:2004:DTP:1037949.1024415} proposed a Miss Ratio Curve (MRC) monitoring hardware feature which can be used as an alternative to page access tracking in the task of WSS estimation~\cite{Zhao:2011:LCW:2002181.2002198}.
\cite{Zhou:2004:DTP:1037949.1024415} proposes a solution which consists in snooping the address bus and requires a collaboration with the OS page fault handler.
As PML/PRL, \cite{Zhou:2004:DTP:1037949.1024415} showed that tracking page accesses at the hardware level is possible.
However, \cite{Zhou:2004:DTP:1037949.1024415} is dedicated to native systems and it needs to collaborate with the OS, unlike PML/PRL.

\paragraph{Checkpointing and live migration}
Extensive research studies have investigated these two topics.
Concerning VM live migration, \cite{Medina:2014:SMM:2578702.2492705,Svard:2015:PPC:2723872.2723894} presented surveys that the reader could refer to.
We would like to highlight among them \cite{2017:LMV:3070823.3070837,Nicolae:2012:HLS:2287076.2287088} who studied live migration of the VM storage along with its memory.
Very few research work has investigated VM storage live migration because it increases the VM downtime.
Migrating the VM storage is necessary when dealing with data intensive applications because they generally use local storage instead of the classical network storage.
The utilization of PRL/PML is also beneficial for this use case because disk accesses go through the buffer cache, which resides in RAM.
\cite{Zhang:2017:MAP:3140607.3050753} addressed another important aspect of live migration which is the prediction of the right migration instant.
In fact, live migration could fail due to lack of resources or sudden VM behavior changes.
The authors use the VM WSS to track such behavior changes.
Thus PRL is likely to improve \cite{Zhang:2017:MAP:3140607.3050753}'s contribution.

Concerning VM checkpointing, we would like to highlight~\cite{Zhang:2013:OVC:2535461.2535463} who focused on optimizing VM restore.
This operation has not been extensively studied for checkpointing although it is of much importance.
In fact, a quick restore reduces the duration of service unavailability after VM failure detection.
The authors minimize disk accesses by optimizing prefectching.
This work is orthogonal to our contribution.

\paragraph{WSS estimation}
{\em Committed\_AS}, a Linux kernel statistic, is generally used (e.g., by Xen) to estimate the VM WSS.
This statistic corresponds to the total number of anonymous memory pages allocated by all processes, but not necessary backed by physical pages.
Therefore, {\em Committed\_AS} over-estimates the WSS.
Another limitation of this approach is the fact that it requires a collaboration between the hypervisor and the guest OS.
Zballoond~\cite{zballoond} relies on the following observation: when a VM's memory size is larger than or equal to its WSS, the number of swap-in and refault (occurs when a previously evicted page is later accessed) events is close to zero.
The basic idea behind {\em Zballoond} consists in gradually decreasing the VM's memory size until these counters start to increase.
The VM's WSS is the lowest memory size which leads the VM to zero swap-in and refault events.
Like {\em Committed\_AS}, {\em Zballoond} requests the collaboration with the guest kernel.
Furthermore, {\em Zballoond} is very active in the sense that it performs memory pressure on the VM, which could degrade the VM performance.
Geiger~\cite{geiger} monitors the evictions and subsequent reloads from the guest OS buffer cache from/to the swap device.
It relies on a ghost buffer~\cite{ghostBuffer} which represents an imaginary memory buffer which extends the VM's physical memory (noted $m_{cur}$).
The size of this buffer (noted $m_{ghost}$) represents the amount of extra memory which would prevent the VM from swapping-out.
Knowing the ghost buffer size, the VM's WSS can be computed using the following formula: $WSS = m_{cur} + m_{ghost}$ if $m_{ghost}>0$.
Unlike the two previous solutions, {\em Geiger} is transparent from the VM's point of view.
However, Geiger has an important drawback which derives from its non-intrusiveness.
It is able to estimate the WSS only when the size of the ghost buffer is greater than zero (the VM is in a swapping state).
Geiger is inefficient if the VM's WSS is smaller than the current memory allocation.
Hypervisor Exclusive Cache~\cite{exclusiveCache} is fairly similar to Geiger.
Badis~\cite{Nitu:2018:WSS:3203302.3179422} combined VMware and {\em Geiger} in order to take advantage of their non intrusivity on the VM's codebase.
Badis suffers from VMware and Geiger's drawbacks presented above.
\cite{Zhao:2011:LCW:2002181.2002198} computes the WSS of an application based on its miss-ratio curve (MRC).
The latter shows the fraction of the cache misses that would turn into cache hits if the VM's allocated memory increases.
\cite{Zhao:2011:LCW:2002181.2002198} presents a set of methods to determine the values of the input parameters of our WSS estimation system ($\tau$, $\omega$, and $\mu$).
\cite{Lee:2011:HAC:2056308.2056420} presents an application-assisted WSS estimation solution in virtualized systems.
In contrast to our solution which considers the VM as a black blox, \cite{Lee:2011:HAC:2056308.2056420} relied on the application inside the VM to estimate the WSS, which is very intrusive.

\section{Conclusion}
\label{conclusion}

This paper presents a thorough analysis of Page Modification Logging (PML), a memory page access tracking technology introduced by Intel and VMware as a key virtualization functionality. 
We show that although the current design of PML makes it effective for VM live migration and checkpointing, it is not appropriate for working set size estimation (WSS).
In the light of our analysis, we propose Page Reference Logging, an extension to PML which makes it also effective for WSS estimation.
We implemented PRL in Gem5, a popular hardware simulator and described a WSS estimation system which leverages PRL.
We evaluated our solution using both real and synthetic applications, and compared it with VMware's WSS estimation solution.
Our results demonstrate that, unlike VMware, our solution is both accurate and does not impact user VMs.


\bibliographystyle{ACM-Reference-Format}
\bibliography{main}

\end{document}